\documentclass[12pt]{article}
\usepackage{amsmath}
\usepackage{amssymb}
\usepackage{amscd}
\usepackage[dvips]{graphicx}
 \linethickness{0.4pt}
\begin{document}
~~
\bigskip
\bigskip
\bigskip
\bigskip
\bigskip
\begin{center}
\section*{{
$\kappa$-DEFORMED STATISTICS AND CLASSICAL FOURMOMENTUM ADDITION
LAW}\footnote{Supported by KBN grant 1P03B01828.}}
\end{center}
\bigskip
\bigskip
\bigskip
\begin{center}
{{\large\bf ${\rm Marcin\;Daszkiewicz}$, ${\rm Jerzy\; Lukierski}$,
${\rm Mariusz\;Woronowicz}$}}
\end{center}
\begin{center}
{
{Institute of Theoretical Physics\\ University of Wroc{\l}aw pl.
Maxa Borna 9, 50-206 Wroc{\l}aw, Poland\\ e-mail:
marcin,lukier,woronow@ift.uni.wroc.pl}}
\end{center}
\bigskip
\bigskip
\bigskip
\bigskip
\bigskip
\bigskip
\begin{abstract}
We consider $\kappa$-deformed relativistic symmetries described
algebraically by modified Majid-Ruegg bicrossproduct basis and
investigate the quantization of field oscillators for the
$\kappa$-deformed free scalar fields on $\kappa$-Minkowski space. By
modification of standard multiplication rule, we postulate the
$\kappa$-deformed algebra of bosonic creation and annihilation
operators.  Our algebra permits to define the n-particle states with
classical addition law for the fourmomenta in a way which is not in
contradiction with the nonsymmetric quantum fourmomentum coproduct.
We introduce $\kappa$-deformed Fock space generated by our
$\kappa$-deformed oscillators which satisfy the standard algebraic
relations with modified $\kappa$-multiplication rule. We show that
such a $\kappa$-deformed bosonic Fock space is endowed with the
conventional bosonic symmetry properties. Finally we discuss the
role of $\kappa$-deformed algebra of oscillators in field-theoretic
noncommutative framework.
\end{abstract}
\bigskip
\bigskip
\bigskip
\bigskip
\eject

\section{Introduction.}

The quantum deformed Poincare symmetries with mass-like deformations
- so-called $\kappa$-deformations \cite{lit1}-\cite{lit5} - were
recently studied as a quantum gravity modifications of the
postulates of classical relativistic symmetries\footnote{In
particular, many results of $\kappa$-deformed symmetry framework
have been adopted by the Doubly Special Relativity formalism (see
e.g. \cite{lit6}-\cite{lit8}).}. If we assume that the
$\kappa$-deformed symmetries are described by the Hopf algebra
scheme there are however problems related mostly with the properties
of coalgebraic sector. In particular, two problems in the framework
of $\kappa$-deformations need to be solved:

i) In all bases the four-momentum coproduct is not symmetric. If we
postulate that such a coproduct describes the total fourmomentum of
multiparticle states, the exchange of particles which does not
modify the total fourmomentum requires new $\kappa$-statistics. In
particular, it should be asked whether one can introduce
consistently the $\kappa$-bosonic multiparticle states with
classical symmetric addition law for the fourmomenta.

ii) Recently, the problem of quantization of noncommutative fields
was studied in the case of canonical noncommutative Minkowski space
coordinates satisfying the relation $[\;{\hat x}_{\mu},{\hat
x}_{\nu}\;] = i\theta_{\mu\nu}$ $(\theta_{\mu\nu} = {\rm const}$;
see e.g. \cite{lit9}-\cite{abe}). In particular it has been shown
\cite{lit10}-\cite{abe} that two modifications - due to
noncommutativity of space-time and due to the
$\theta_{\mu\nu}$-statistics - can cancel each other\footnote{Such a
cancelation implies that the Green functions and subsequently
S-matrix remain $\theta_{\mu\nu}$-independent. In \cite{lit11},
\cite{abe} there was presented also the general case with two
independent $\theta$-parameters describing respectively the formula
for star product and for the deformed oscillator algebra. The case
leading to the cancelation is justified \cite{lit10} by the
covariance under the twisted Poincare group.}. It is interesting to
determine the interplay of the star multiplication of space-time
fields and deformed statistics in the case of $\kappa$-deformed
quantum fields.

In this paper we shall resolve the first problem, and provide the
tools for solving the second problem - the $\kappa$-deformed algebra
of field oscillators and consistent braided relations for two copies
of $\kappa$-Minkowski space-time coordinates.

We shall show that solving the fourmomentum addition problem leads
to new $\kappa$-quantization rules for the creation and annihilation
operators, which imply that the change of order of oscillators
produces the modification of their threemomentum dependence. Such a
novel feature does not occur in the case of canonical
$\theta_{\mu\nu}$-deformation, however if $\theta_{\mu\nu}$ depends
on $x_{\mu}$ in linear way (i.e. in momentum space $\theta_{\mu\nu}$
depends linearly on fourmomentum derivative), there is generated a
shift in the momentum dependence of exchanged field oscillators.

In our discussion we shall use the quantum $\kappa$-deformed
Poincare symmetries formulated in modified bicrossproduct basis with
classical Lorentz subalgebra, and the $\kappa$-deformed mass-shell
invariant under the change $P_0 \to -P_0$ \cite{lit12a}. Such a
choice is obtained if in standard bicross-product basis \cite{lit4},
\cite{lit5} we change $P_i \to {\rm e}^{\frac{P_0}{2\kappa}}P_i$,
i.e. if we keep the fourmomenta the same as in the standard
"historical" basis \cite{lit1}-\cite{lit3}. Then our
$\kappa$-deformed Poincar\'{e} algebra with generators $M_{\mu\nu} =
(M_i = \frac{1}{2}\epsilon_{ijk}M_{jk}, N_i = M_{i0})$ and $P_{\mu}
= (P_i,P_0)$ satisfy the following Hopf algebra relations:

a) algebraic sector
\begin{equation}
[\;M_{\mu\nu},M_{\lambda\sigma}\;] = i\left(
\eta_{\mu\sigma}M_{\nu\lambda} - \eta_{\nu\sigma}M_{\mu\lambda} +
\eta_{\nu\lambda}M_{\mu\sigma} - \eta_{\mu\lambda}M_{\nu\sigma}
\right)\;,\;\;\;\;~~\;\;\;\;~~\;\;\;\;~~\; \label{a1}
\end{equation}
\begin{equation}
[\;M_i,P_j\;] = i\epsilon_{ijk}P_k
\;,\;\;\;\;\;\;\;\;\;~~\;\;\;\;~~\;\;\;\;~~\;\;\;
\;\;\;\;~~\;\;\;\;~~\;\;\;\;~~\;\;\;\;\;\;~~\;\;\;\;~~\;\;\;\;~~\;\;\label{a2}
\end{equation}
\begin{equation}
[\;N_{i},P_j\;] = i\delta_{ij} {\rm
e}^{\frac{P_0}{2\kappa}}\left[\frac{\kappa}{2}\left(1-{\rm
e}^{-\frac{2P_0}{\kappa}}\right) + \frac{1}{2\kappa}{\rm
e}^{-\frac{P_0}{\kappa}} \vec{P}^2 \right] - \frac{i}{2\kappa}{\rm
e}^{-\frac{P_0}{2\kappa}}P_iP_j \;, \label{a3}
\end{equation}
\begin{equation}
[\;M_{i},P_{0}\;] = 0 \;\;,\;\;\;\;[\;N_{i},P_{0}\;] = i{\rm
e}^{-\frac{P_0}{2\kappa}}P_i\;,
~~\;\;\;\;\;\;~~\;\;\;\;\;\;~~\;\;\;\;\;\;~~\;\;\;\;\;\;~~\;\;\;\;\;\label{a4}
\end{equation}
\begin{equation}
[\;P_{\mu},P_{\nu}\;] =
0\;,\;~~\;\;\;\;\;\;~~\;\;\;\;\;\;~~\;\;\;\;\;~~\;\;\;\;\;\;~~\;\;\;\;\;\;~~\;\;\;\;\;~~
\;\;\;\;\;\;~~\;\;\;\;\;\;~~\;\;\;~~~~~\label{a5}
\end{equation}

b) coalgebra sector
\begin{equation}
\Delta(M_i) = M_i\otimes 1 + 1\otimes
M_i\;,\;\;\;\;~~\;\;\;\;\;\;\;~~\;\;\;\;\;\;\;~~\;\;\;\;\;\;\;~\;\;
\label{c1}
\end{equation}
\begin{equation}
 \Delta(N_i) =
N_i\otimes 1 + {\rm e}^{-\frac{P_0}{\kappa}}\otimes N_i +
\frac{1}{\kappa}\epsilon_{ijk}{\rm
e}^{-\frac{P_0}{2\kappa}}P_j\otimes M_k\;,\label{c2}
\end{equation}
\begin{equation}
\Delta(P_0) = P_0\otimes 1 + 1\otimes
P_0\;,\;\;\;\;~~\;\;\;\;\;\;\;~~\;\;\;\;\;\;\;~~\;\;\;\;\;\;\;\;\;\;~~\label{c3}
\end{equation}
\begin{equation}
 \Delta(P_i) =
P_i\otimes {\rm e}^{\frac{P_0}{2\kappa}} + {\rm
e}^{-\frac{P_0}{2\kappa}}\otimes P_i\;,
\;\;\;\;~~\;\;\;\;\;\;\;~~\;\;\;\;\;\;\;\;\;\;~~\;\label{c4}
\end{equation}

c) antipodes
\begin{equation}
S(M_i) = -M_i\;\;,\;\;S(N_i) = - {\rm e}^{\frac{P_0}{\kappa}}N_i
+\frac{1}{\kappa} \epsilon_{ijk}{\rm e}^{\frac{P_0}{2\kappa}}P_jM_k
\;,\label{anti1}
\end{equation}
\begin{equation}
S(P_i) = -P_i\;\;,\;\;S(P_0) = -P_0\;. \label{anti2}
\end{equation}
The $\kappa$-Poincare algebra has two $\kappa$-deformed Casimirs
describing mass and spin. The deformation of bilinear mass Casimir
looks as follows\footnote{This Casimir is identical with the one in
so-called "standard basis" \cite{lit1}, with $\kappa$-deformed
boosts algebra.}
\begin{equation}
C_2^{\kappa} (P_\mu) = C_2^{\kappa} (\vec{P}^2,P_0) = \left
(2\kappa\sinh \left(\frac{P_0}{2\kappa}\right)\right)^2 -
\vec{P}^2\;. \label{casimir}
\end{equation}
Because in any basis $C_2^{\kappa} (P_\mu) = C_2^{\kappa}
(S(P_\mu))$ we see that for the choice (\ref{casimir}) any solution
with $P_0>0$ has its "antiparticle counterpart" with $P_0 \to -P_0$.

The $\kappa$-deformed Minkowski space is defined as the translation
sector of dual $\kappa$-Poincar\'{e} group \cite{lit4}, \cite{lit12}
by the following relations
\begin{equation}
[\;{\hat x}_{0},{\hat x}_{i}\;] = \frac{i}{\kappa}{\hat
x}_{i}\;\;,\;\; [\;{\hat x}_{i},{\hat x}_{j}\;] = 0\;.
\label{minkowski}
\end{equation}
In this paper we propose how to introduce the $\kappa$-statistics
defining $\kappa$-bosons\footnote{We shall discuss only bosonic
case; the $\kappa$-fermions can be introduced by adding obvious sign
factors.} and provide a consistent description of quantum bosonic
fields defined on $\kappa$-deformed Minkowski space
(\ref{minkowski}). In Sect. 2 we shall consider the noncommutative
free scalar fields with field quanta described by $\kappa$-deformed
algebra of creation and annihilation operators, with deformation
encoded in deformed multiplication rule. We shall consider the
noncommutative Fourier transform using the following
$\kappa$-deformation of plane waves \cite{lit12b}, \cite{lit12a}
\begin{equation}
\vdots {\rm e}^{ip_\mu \hat{x}^\mu}\vdots ={\rm
e}^{\frac{i}{2}p_0{\hat x}^{0}}{\rm e}^{ip_i{\hat x}^{i}} {\rm
e}^{\frac{i}{2}p_0{\hat x}^{0}} ={\rm e}^{i{\tilde p}_i{\hat
x}^{i}}{\rm e}^{ip_0{\hat x}^{0}} \;, \label{exp}
\end{equation}
where ${\tilde p}_i\equiv{\tilde p}_i(\vec{p},p_0) = {\rm
e}^{-\frac{p_0}{2\kappa}} p_i\;. \label{p}$ \\
From (\ref{exp}) follows the simple Hermitean conjugation property
\begin{equation}
\left(\vdots {\rm e}^{ip_\mu \hat{x}^\mu}\vdots \right)^\dag =
\left({\rm e}^{i{\tilde p}_i{\hat x}^{i}}{\rm e}^{ip_0{\hat
x}^{0}}\right)^{\dag} = {\rm e}^{-ip_0{\hat x}^{0}}{\rm
e}^{-i{\tilde{\tilde p}}_i{\hat x}^{i}}=\vdots {\rm e}^{-ip_\mu
\hat{x}^\mu}\vdots \;, \label{property}
\end{equation}
with ${\tilde{\tilde p}}_i = {{\tilde p}}_i(-p_0,\vec{p}) = {\rm
e}^{\frac{p_0}{2\kappa}} p_i$. We also observe (see (\ref{anti2}))
that $S({\tilde p}_i) = -{\tilde{\tilde p}}_i$.

Using the expansion of noncommutative free field into the
$\kappa$-deformed plane waves (\ref{exp}) we introduce the
creation/annihilation operators satisfying the $\kappa$-deformed
field oscillators algebra. The novel feature of our construction is the 
new multiplication rule of creation/anihilation operators which
requires putting them off-shell. We shall introduce the algebraic
properties of these operators in such a way that the difficulty with
the interpretation of nonsymmetric quantum addition law for the
fourmomenta (see (\ref{c3}), (\ref{c4})) will be removed. The
$\kappa$-deformation of the oscillator algebra  described by a new
multiplication rule $(a\cdot b \to a\circ b)$,  will be also used in
Sect. 3 for the construction of $\kappa$-deformed Fock space.
Surprisingly, the Fock space constructed within the
$\kappa$-deformed Hopf algebraic scheme will have similar structure
as the classical one with on-shell fourmomenta, and with symmetric
multiparticle states and classical fourmomentum addition law.
Finally in Sect. 4 we shall propose a braided structure of tensor
product of two $\kappa$-Minkowski spaces as a first step in the
construction of quantized $\kappa$-deformed local field theory.

It should be pointed out that in this paper we formulated the new
$\kappa$-quantization rules for the creation and annihilation
operators (modulo possibly some numerical normalization factors).
The derivation of the respective noncommutative structure of quantum
free fields in $\kappa$-Minkowski space is presented in our next
publication \cite{our}. In particular, because the $\kappa$-deformed
K-G operator in commuting fourmomentum space is not a polynomial,
the standard canonical quantization scheme with E.T. commutators has
to be essentially modified (see \cite{our}), but if we use suitably
chosen kappa-multiplication of fields, the field quantization rules
can be put back into canonical form.

\section{The quantization of $\kappa$-deformed free K-G fields.}

We describe the $\kappa$-deformed scalar free field on
noncommutative Minkowski space (\ref{minkowski}) by decomposition
into field quanta creation and annihilation operators, with the use
of $\kappa$-deformed Fourier transform (see (\ref{exp}))
\begin{equation}
\phi ({\hat x}) = \frac{1}{(2\pi)^{3/2}} \int d^4p\;
A(p_0,\vec{p})\;\delta \left(C_2^{\kappa} (\vec{p}^2,p_0) -
M^2\right)\vdots {\rm e}^{ip_\mu \hat{x}^\mu}\vdots\;. \label{field}
\end{equation}
It should be stressed that the $\kappa$-deformed bicovariant
calculus implies (see \cite{lit16}, \cite{lit17}), that the free
fields satisfying noncommutative K-G equation have the
$\kappa$-deformed mass-shell modified as follows
\begin{equation}
\delta\left(C_2^{\kappa} - M^2\right) \to \delta\left [
C_2^{\kappa}\left (1 - \frac{C_2^{\kappa}}{4\kappa^2}\right ) - M^2
\right ]\;. \label{ms}
\end{equation}
The corresponding free field can be described by a superposition of
two fields of type (\ref{field}) - one physical and the second one
describing ghosts \cite{lit18}, \cite{lit19}.

The mass-shell condition in formula (\ref{field}) can be solved and
leads to the following $\kappa$-deformed energy-momentum dispersion
relation (see e.g. \cite{lit5})
\begin{equation}
p_0^{\pm} =
\pm\omega_\kappa(\vec{p}^2)\;\;,\;\;\omega_\kappa(\vec{p}^2) =
2\kappa~{\rm arcsinh}\left (\frac{\sqrt{\vec{p}^2 +M^2}}{2\kappa}
\right)\;. \label{dis}
\end{equation}
The free field (\ref{field}) can be decomposed into positive and
negative frequency parts
\begin{equation}
\phi({\hat x}) = \phi_{+}({\hat x}) + \phi_{-}({\hat x})\;.
\label{decom}
\end{equation}
The negative frequency part $\left( p_0^- =
-\omega_\kappa(\vec{p}^2) \right)$ provides for the complex field
$\phi({\hat x})$ the annihilation operators for the antiparticles.
\\
Using the relation
\begin{equation}
\delta \left(C_2^{\kappa} (\vec{p}^2,p_0) - M^2\right)  =
\frac{1}{\Omega_{+}(\vec{p}^2)}\delta(p_0 - p_0^{+}) +
\frac{1}{\Omega_{-}(\vec{p}^2)}\delta(p_0 - p_0^{-})\;, \label{rel}
\end{equation}
where
\begin{equation}
{\Omega_{\pm}(\vec{p}^2)} = \left|\frac{\partial}{\partial
p_0}C_2^{\kappa} (\vec{p}^2,p_0) \right|_{p_0 = p_0^{\pm}} = 2\kappa
\sinh\left(\frac{\omega_{\kappa}(\vec{p}^2)}{\kappa}\right) \equiv
\Omega_{\kappa} (\vec{p}^2)\;, \label{omega}
\end{equation}
we obtain
\begin{equation}
\phi_{\pm} ({\hat x}) = \frac{1}{(2\pi)^{3/2}} \int
\frac{d^3\vec{p}}{\Omega_{\kappa} (\vec{p}^2)}\;
A(\pm\omega_{\kappa}(\vec{p}^2),\pm \vec{p})\;  \vdots {\rm e}^{\pm
ip_\mu \hat{x}^\mu}\vdots |_{p_0 = \omega_{\kappa}(\vec{p}^2)}\;.
\label{fieldpm}
\end{equation}
In order to obtain the $\kappa$-deformed real scalar field (i.e. we
identify the particles and antiparticles)
\begin{equation}
\left( \phi_{\pm}({\hat x})\right)^\dag = \phi_{\mp}({\hat x}) \;,
\label{real}
\end{equation}
one should assume that\footnote{The simplicity of the relation
(\ref{ass}) follows from the relations (\ref{anti2}) and
(\ref{property}); in arbitrary fourmomentum basis we have
\begin{equation*}
\left(A(p_0^+,\vec{p}) \right)^\dag = A(S(p_0^+),S(\vec{p}))
\nonumber \;.
\end{equation*}}
\begin{equation}
\left(A(\pm\omega_{\kappa}(\vec{p}^2),\pm \vec{p}) \right)^\dag =
A(\mp\omega_{\kappa}(\vec{p}^2),\mp \vec{p}) \;. \label{ass}
\end{equation}

One can introduce the on-shell creation and annihilation operators
by means of the relations
\begin{eqnarray}
a_\kappa (\omega_{\kappa}(\vec{p}^2),\vec{p}) =
C(\vec{p}^2)A(\omega_{\kappa}(\vec{p}^2),\vec{p})\;\;,\;\;
a_{\kappa}^\dag (\omega_{\kappa}(\vec{p}^2),\vec{p}) =
C(\vec{p}^2)A(-\omega_{\kappa}(\vec{p}^2),-\vec{p}) \;,
\label{creation}
\end{eqnarray}
where as for the normalization factor we choose $C(\vec{p}^2) =
\Omega_{\kappa}^{\frac{1}{2}} (\vec{p}^2)$. \\
The $\kappa$-statistics  will be introduced in a way consistent with
the addition law (\ref{c4}) of fourmomenta. It was already pointed
out (see e.g. \cite{lit20}, Sect. 5), that the standard
(anti)commutativity of the operators (\ref{creation}) is not
consistent with the quantum composition law for the fourmomenta
\begin{equation}
\vec{p}_{1+2}  = \vec{p}_1{\rm e}^{\frac{p_{02}}{2\kappa}} + {\rm
e}^{-\frac{p_{01}}{2\kappa}}\vec{p}_2  \neq \vec{p}_{2+1} =
\vec{p}_2{\rm e}^{\frac{p_{01}}{2\kappa}} + {\rm
e}^{-\frac{p_{02}}{2\kappa}} \vec{p}_1 \;. \label{law}
\end{equation}
\\
We shall use the algebraic relations expressing the property that
the operators (\ref{creation}) create/annihilate quanta with
fourmomentum $p_{\mu}$
\begin{equation}
P_{\mu} \triangleright a_\kappa (p_0,\vec{p}) = p_\mu a_\kappa
(p_0,\vec{p})\;\;,\;\; P_{\mu} \triangleright a_\kappa^\dag
(p_0,\vec{p}) = -p_\mu a_\kappa^\dag (p_0,\vec{p})\;, \label{arel}
\end{equation}
as well as\footnote{Analogous deformed Leibnitz rule relations can
be written for $a^\dag a^\dag$, $a a^\dag$ and $a^\dag a$ operator
products.}
\begin{equation}
P_{\mu} \triangleright  \left( a_\kappa (p_0,\vec{p})\;a_\kappa
(q_0,\vec{q}) \right) = \left( \Delta^{(1)}(P_\mu) \triangleright
a_\kappa (p_0,\vec{p}) \right) \left( \Delta^{(2)}(P_\mu)
\triangleright a_\kappa (q_0,\vec{q}) \right) \;,\label{arel3}
\vspace{0.1cm}
\end{equation}
where $\Delta(P_\mu) \equiv \Delta^{(1)}(P_\mu)\otimes
\Delta^{(2)}(P_\mu)$ is a shorthand notation of coproducts
(\ref{c3}), (\ref{c4}).

In order to present our construction we assume that the relations
(\ref{arel}), (\ref{arel3}) can be extended also for $p_0$ not
necessarily equal to $\omega_{\kappa}(\vec{p})$. In such a way we
introduce the off-shell field oscillators, which will be useful for
our construction.

The general $\kappa$-deformed creation operators algebra defining
the $\kappa$-statistics can be written as follows\footnote{We
consider the most general relation in a given fourmomentum frame
with fixed coproducts see ((\ref{c3}), (\ref{c4})), i.e. we are not
allowed to simplify (\ref{nccr1}) by nonlinear transformations of
the arguments of $a_{\kappa}$.}
\begin{equation*}
F_{\kappa}(p,q)a_\kappa (f_0(p,q),\vec{f}(p,q)) a_\kappa
\left(g_0(p,q),\vec{g}(p,q)\right)
=\;\;\;\;\;\;\;\;\;\;\;\;\;\;\;\;\;\;\;\;\;\;\;\;\;\;\;\;\;\;\;\;\;\;\;\;\;\;\;\;\;\;\;\;\;\;\;
\end{equation*}
\begin{equation}
\;\;\;\;\;\;\;\;\;\;\;\;\;\;\;\;\;\;\;\;\;\;\;\;\;\;\;\;\;\;\;\;\;\;\;\;\;\;\;\;\;\;\;\;\;\;\;=G_{\kappa}(q,p)a_\kappa
(h_0(p,q),\vec{h}(p,q)) a_\kappa (k_0(p,q),\vec{k}(p,q))
\;,\label{nccr1}
\end{equation}
where the functions $F_{\kappa}$ and $G_{\kappa}$ are suitably
restricted by multioscillator algebraic relations. Further we should
assume that $\lim\limits_{\kappa \to \infty}F_{\kappa}(p,q) =
\lim\limits_{\kappa \to \infty}G_{\kappa}(p,q) = 1$ and
\begin{eqnarray}
&&\lim_{\kappa \to \infty} f_0 = p_0\;\;\;,\;\;\;\lim_{\kappa \to
\infty} g_0 = q_0\;\;\;,\;\;\;\lim_{\kappa \to \infty} h_0 =
q_0\;\;\;,\;\;\; \lim_{\kappa \to \infty} k_0 = p_0\;,
\label{limes1}\\
&&\lim_{\kappa \to \infty} f_i = p_i\;\;\;,\;\;\;\;\lim_{\kappa \to
\infty} g_i = q_i\;\;\;,\;\;\;\;\lim_{\kappa \to \infty} h_i =
q_i\;\;\;,\;\;\;\;\lim_{\kappa \to \infty} k_i = p_i\;. \nonumber
\end{eqnarray}
The consistent action of the fourmomentum operator on  both sides of
the eq. (\ref{nccr1}), from (\ref{arel}), (\ref{arel3}), leads to
the following relations
\begin{equation}
f_0 + g_0 = h_0 + k_0\;\;\;,\;\;\; \vec{f}{\rm
e}^{\frac{g_0}{2\kappa}} + \vec{g}{\rm e}^{-\frac{f_0}{2\kappa}} =
\vec{h}{\rm e}^{\frac{k_0}{2\kappa}} + \vec{k}{\rm
e}^{-\frac{h_0}{2\kappa}}\;.
 \label{relations1}
\end{equation}
We shall further assume that the relations (\ref{relations1})
describe classical addition of fourmomenta, what can be achieved if
we choose
\begin{equation}
f_0 = p_0\;\;\;,\;\;\;g_0 = q_0\;\;\;,\;\;\;h_0 =
q_0\;\;\;,\;\;\;k_0 = p_0\;, \label{cllas1}
\end{equation}
\begin{equation}
f_i = p_i{\rm e}^{-\frac{q_0}{2\kappa}}\;\;\;,\;\;\;g_i = q_i{\rm
e}^{\frac{p_0}{2\kappa}}\;\;\;,\;\;\;h_i = q_i{\rm
e}^{-\frac{p_0}{2\kappa}}\;\;\;,\;\;\;k_i = p_i{\rm
e}^{\frac{q_0}{2\kappa}}\;. \label{cllas2}
\end{equation}
The relation (\ref{nccr1}) takes the form\footnote{Further we shall
consider the case with $F_{\kappa}(p,q) = 1$ and $G_{\kappa}(q,p) =
1$.}
\begin{equation}
a_\kappa \left(p_0,{\rm e}^{-\frac{q_0}{2\kappa}}\vec{p} \right)
a_\kappa \left(q_0,{\rm e}^{\frac{p_0}{2\kappa}}\vec{q} \right) =
a_\kappa \left(q_0,{\rm e}^{-\frac{p_0}{2\kappa}}\vec{q} \right)
a_\kappa \left(p_0,{\rm e}^{\frac{q_0}{2\kappa}}\vec{p}\right)
\;,\label{ccr1}
\end{equation}
where we should put $p_0 = \omega_{\kappa}(\vec{p}^2)$ and $q_0 =
\omega_{\kappa}(\vec{q}^2)$. We see from (\ref{ccr1}) that the
fourvectors $(p_0,{\rm e}^{\pm\frac{q_0}{2\kappa}}\vec{p})$ and
$(q_0,{\rm e}^{\pm\frac{p_0}{2\kappa}}\vec{q})$ are off-shell, but
this effect occurs only if we consider the product of oscillators.
It will appear however that due to such a construction one can
define the $\kappa$-deformed 2-particle states with Abelian
fourmomentum addition law $(\vec{p}+ \vec{q},
\omega_{\kappa}(\vec{p})+ \omega_{\kappa}(\vec{q}))$ for two
on-shell fourmomenta $(\vec{p}, \omega_{\kappa}(\vec{p}))$,
$(\vec{q}, \omega_{\kappa}(\vec{q}))$.

It should be noted that if we change the order of two creation
operators in (\ref{ccr1}), their threemomenta dependence is modified
in a way depending on the energy of second partner. We claim that
this "knowledge" of one particle about the energies of other
particles is the geometric effect induced by the noncanonical
(Lie-algebraic) noncommutativity structure of $\kappa$-Minkowski
space-time.

Using the conjugation property (\ref{ass}) and the definitions
(\ref{creation}) one gets from (\ref{ccr1}) the relation for the
annihilation operators
\begin{equation}
a_\kappa^\dag \left(p_0,{\rm e}^{\frac{q_0}{2\kappa}}\vec{p} \right)
a_\kappa^\dag \left(q_0,{\rm e}^{-\frac{p_0}{2\kappa}}\vec{q}
\right) = a_\kappa^\dag \left(q_0,{\rm
e}^{\frac{p_0}{2\kappa}}\vec{q} \right) a_\kappa^\dag \left(p_0,{\rm
e}^{-\frac{q_0}{2\kappa}}\vec{p}\right) \;.\label{ccr2}
\end{equation}
The remaining algebraic relation, which is consistent with the
Hermitean conjugation (\ref{ass}) and Abelian addition law for
fourmomenta looks as follows (we recall that $p_0 =
\omega_{\kappa}(\vec{p}^2)$, $q_0 = \omega_{\kappa}(\vec{q}^2)$)
\begin{equation}
a_\kappa^\dag \left(p_0,{\rm e}^{-\frac{q_0}{2\kappa}}\vec{p}
\right) a_\kappa \left(q_0,{\rm e}^{-\frac{p_0}{2\kappa}}\vec{q}
\right) - a_\kappa \left(q_0,{\rm e}^{\frac{p_0}{2\kappa}}\vec{q}
\right) a_\kappa^\dag \left(p_0,{\rm
e}^{\frac{q_0}{2\kappa}}\vec{p}\right) = \delta^{(3)}\left(\vec{p} -
\vec{q} \right)\;.\label{ccr3}
\end{equation}

The $\kappa$-deformed oscillator algebra (\ref{ccr1}-\ref{ccr3}) can
be rewritten in known classical form if we define the following
$\kappa$-multiplication of the oscillators $a _\kappa(p )$
\begin{equation}
a_\kappa (p)\circ a_\kappa(q):=a_\kappa \left(p_0,{\rm
e}^{-\frac{q_0}{2\kappa}}\vec{p} \right) a_\kappa \left(q_0,{\rm
e}^{\frac{p_0}{2\kappa}}\vec{q} \right)\;, \label{multi}
\end{equation}
and when, in consistency with the conjugation relation
$a_{\kappa}^\dag (p) = a_{\kappa}(-p)$ (see (\ref{ass}),
(\ref{creation})), we introduce
\begin{equation}
a_\kappa^\dag (p)\circ a_\kappa^\dag (q)= a_\kappa^\dag
\left(p_0,{\rm e}^{\frac{q_0}{2\kappa}}\vec{p} \right) a_\kappa^\dag
\left(q_0,{\rm e}^{-\frac{p_0}{2\kappa}}\vec{q} \right)\;,\;\;\;\;\;
\label{multi1}
\end{equation}

\begin{equation}
a_\kappa^\dag (p)\circ a_\kappa(q)= a_\kappa^\dag \left(p_0,{\rm
e}^{-\frac{q_0}{2\kappa}}\vec{p} \right) a_\kappa \left(q_0,{\rm
e}^{-\frac{p_0}{2\kappa}}\vec{q} \right)\;,\;\; \label{multi2}
\end{equation}
\vspace{0.03cm}
\begin{equation}
a_\kappa (p)\circ a_\kappa^\dag(q)= a_\kappa\left(p_0,{\rm
e}^{\frac{q_0}{2\kappa}}\vec{p} \right) a_\kappa^\dag \left(q_0,{\rm
e}^{\frac{p_0}{2\kappa}}\vec{q} \right)\;.
\;\;\;\;\;\;~\label{multi3}
\end{equation}
We see that in place of the relations (\ref{ccr1})-(\ref{ccr3}) we
obtain the standard algebra of creation and annihilation operators
$([\;A,B\;]_\circ := A\circ B - B\circ A)$
\begin{equation}
[\;a_\kappa  (p),a_\kappa  (q)\;]_{\circ} =  [\;a_\kappa ^\dag
(p),a_\kappa ^\dag (q)\;]_{\circ} = 0\;\;,\;\;[\;a_\kappa ^\dag
(p),a_\kappa  (q)\;]_{\circ} = \delta^{(3)} (\vec{p}-\vec{q})\;.
\label{stanccr}
\end{equation}
In order to extend the multiplication (\ref{multi}) to the product
of any oscillators we define the following triple $\kappa$-product
\begin{equation}
(a_\kappa (p)\circ a_\kappa (q))\circ a_\kappa (r) =
a_\kappa\left(p_0,{\rm e}^{-\frac{(q_0+r_0)}{2\kappa}}\vec{p}
\right) a_\kappa\left(q_0,{\rm e}^{\frac{(p_0-r_0)}{2\kappa}}\vec{q}
\right)a_\kappa\left(r_0,{\rm e}^{\frac{(p_0+q_0)}{2\kappa}}\vec{r}
\right)\;. \label{triple}
\end{equation}
It is easy to check that
\begin{equation}
(a_\kappa  (p)\circ a_\kappa (q))\circ a_\kappa (r)  = a_\kappa
(p)\circ (a_\kappa (q)\circ a_\kappa (r))\;,\label{aso}
\end{equation}
i.e. the $\kappa$-multiplication of oscillators is associative.\\
The associative $\kappa$-deformed product of n oscillators looks as
follows
\begin{equation}
~\;\;\;\;\;\;\;\;\;\;a_\kappa(p^{(1)})\circ\ldots \circ
a_\kappa(p^{(k)})\circ\ldots \circ
a_\kappa(p^{(n)})=a_\kappa\left(p_0^{(1)},{\rm
exp}\left(-\sum_{i=2}^{n}\frac{p_0^{(i)}}{2\kappa}
\right)\vec{p}^{~(1)} \right) \ldots
\;\;\;\;\;\;\;\;\;\;\;\;\;\;\;\;\;\; \label{hakk}
\end{equation}
\begin{equation*}
\ldots a_\kappa\left(p_0^{(k)},{\rm exp
}\left(-\sum_{i=k+1}^{n}\frac{p_0^{(i)}}{2\kappa} +
\sum_{i=1}^{k-1}\frac{p_0^{(i)}}{2\kappa} \right)\vec{p}^{~(k)}
\right)\ldots  a_\kappa\left(p_0^{(n)},{\rm
exp}\left(\sum_{i=1}^{n-1}\frac{p_0^{(i)}}{2\kappa}
\right)\vec{p}^{~(n)} \right)\;.\;\;\;\;\;\;\;\;\;\;\;\;\;\;\;
\nonumber
\end{equation*}
Finally, we extend the formula (\ref{hakk}) to arbitrary normally
ordered product of creation and annihilation operators
\begin{equation}
a^\dag_\kappa(k^{(1)})\circ\ldots \circ a^\dag_\kappa(k^{(m)})\circ
a_\kappa(p^{(1)})\circ\ldots \circ a_\kappa(p^{(n)})=
\end{equation}
\begin{equation*}
a^\dag_\kappa\left(k_0^{(1)},{\rm exp
}\left(\sum_{i=2}^{m}\frac{k_0^{(i)}}{2\kappa} -
\sum_{i=1}^{n}\frac{p_0^{(i)}}{2\kappa} \right)\vec{k}^{(1)}
\right)\ldots a^\dag_\kappa\left(k_0^{(m)},{\rm exp
}\left(\sum_{i=1}^{m-1}\frac{k_0^{(i)}}{2\kappa} -
\sum_{i=1}^{n}\frac{p_0^{(i)}}{2\kappa} \right)\vec{k}^{(m)} \right)
\end{equation*}
\begin{equation*}
a_\kappa\left(p_0^{(1)},{\rm exp
}\left(-\sum_{i=1}^{m}\frac{k_0^{(i)}}{2\kappa} -
\sum_{i=2}^{n}\frac{p_0^{(i)}}{2\kappa} \right)\vec{p}^{~(1)}
\right)\ldots a_\kappa\left(p_0^{(n)},{\rm exp
}\left(-\sum_{i=1}^{m}\frac{k_0^{(i)}}{2\kappa} +
\sum_{i=1}^{n-1}\frac{p_0^{(i)}}{2\kappa} \right)\vec{p}^{~(n)}
\right)
\end{equation*}

\section{$\kappa$-deformed Fock space for $\kappa$-bosons.}

Let us introduce the normalized vacuum state in standard way
\begin{equation}
a_\kappa^\dag (p_0,\vec{p})|0> = 0\;\;\;,\;\;\; <0|0> = 1\;,
\label{vacuum}
\end{equation}
where $p_0 = \omega_{\kappa}(\vec{p}^2)$. The one-particle state is
defined as follows
\begin{equation}
|\vec{p}> =  a_\kappa(p_0,\vec{p})|0>\;, \label{one}
\end{equation}
where from (\ref{arel}) we get
\begin{equation}
P_{\mu}|\vec{p}> =  p_{\mu}|\vec{p}>\;. \label{pone}
\end{equation}
If we define two-particle states in the following way
\begin{equation}
|\vec{p},\vec{q}> =  a_\kappa(p_0,\vec{p})\circ
a_\kappa(q_0,\vec{q})|0>\;, \label{two}
\end{equation}
by using (\ref{arel3}) and (\ref{multi}) we get
\begin{equation}
P_{\mu}|\vec{p},\vec{q}> =  (p_{\mu}+q_{\mu})|\vec{p},\vec{q}>\;,
\label{ptwo}
\end{equation}
where we see from (\ref{two}) that fourmomenta $(p_\mu,q_\mu)$ are
on-shell.  From (\ref{stanccr}) follows the classical bosonic
symmetry
\begin{equation}
|\vec{p},\vec{q}> = |\vec{q},\vec{p}>\;. \label{symmetryb}
\end{equation}
Due to the associativity property  of $\kappa$-multiplication
(\ref{aso}) one can define the n-particle state by the
$\kappa$-deformed product of n oscillators. We obtain
\begin{equation}
|\vec{p}^{~(1)},\ldots,\vec{p}^{~(k)},\ldots,\vec{p}^{~(n)}> =
a_\kappa(p^{(1)})\circ \ldots \circ a_\kappa(p^{(k)})\circ \ldots
\circ a_\kappa(p^{(n)})|0> \label{state}\,.
\end{equation}
Extending by iteration the coassociative coproduct (\ref{c3}) and
(\ref{c4}) to the tensor product of n-th order we can calculate the
total momentum of the state (\ref{state}), which appears to be given
by classical formula
\begin{equation*}
P_\mu |\vec{p}^{~(1)},\ldots,\vec{p}^{~(k)},\ldots,\vec{p}^{~(n)}> =
\left [ \Delta^{(n)}(P_\mu) \triangleright
\left(a_\kappa(p^{(1)})\circ \ldots \circ a_\kappa(p^{(n)})
\right)\right]|0> = \;\;\;\;\;\;\;\;\;\;\;\nonumber
\end{equation*}
\begin{equation}
~\;\;\;\;\;\;\;\;\;\;\;\;\;\;\;\;\;\;\;\;\;\;\;\;\;\;\;\;\;\;\;\;\;\;\;\;\;\;\;\;\;\;
=\sum_{i=1}^{n}p_{\mu}^{(i)}
|\vec{p}^{~(1)},\ldots,\vec{p}^{~(k)},\ldots,\vec{p}^{~(n)}>\;,\label{pstate}
\end{equation}
where again the fourmomenta $p_\mu^{(i)}$ are on-shell. We get
immediately from (\ref{stanccr}) also the classical bosonic symmetry
property
\begin{equation}
|\vec{p}^{~(1)},\ldots,\vec{p}^{~(i)},\ldots,\vec{p}^{~(j)},\ldots,\vec{p}^{~(n)}>
=
|\vec{p}^{~(1)},\ldots,\vec{p}^{~(j)},\ldots,\vec{p}^{~(i)},\ldots,\vec{p}^{~(n)}>\;.
\label{pstaterrr}
\end{equation}

In order to complete the structure of $\kappa$-deformed Fock space
we should define dual vectors and scalar product. We define the dual
space in analogy to the relations (\ref{state})
\begin{equation}
<\vec{k}^{(1)},\ldots,\vec{k}^{(n)}| = <0|a_\kappa^\dag
(k_0^{(1)},\vec{k}^{(1)})\circ \ldots \circ a_\kappa^\dag
(k_0^{(n)},\vec{k}^{(n)})\;. \label{dual}
\end{equation}
We define $\kappa$-deformed scalar product of the basic vectors
(\ref{state}) and (\ref{dual}) as follows
\begin{equation*}
~\;\;\;\;\;\;\;\;<\vec{k}^{(1)},\ldots,\vec{k}^{(m)}|\vec{p}^{~(1)},\ldots,\vec{p}^{~(n)}>_{\kappa}
:=\; <\vec{k}^{(1)},\ldots,\vec{k}^{(m)}|\circ
|\vec{p}^{~(1)},\ldots,\vec{p}^{~(n)}> = \;\;\;\;\;\;\;\;\;\;
\;\;\;\;\;\;\;\;\;\;\;\;\;\;\;\;\;\;\;\;\nonumber
\end{equation*}
\begin{equation}
=<0|a_\kappa^\dag (k_0^{(1)},\vec{k}^{(1)})\circ \ldots \circ
a_\kappa^\dag (k_0^{(m)},\vec{k}^{(m)})\circ
a_\kappa(p_0^{(1)},\vec{p}^{~(1)})\circ \ldots \circ
a_\kappa(p_0^{(n)},\vec{p}^{~(n)})|0>\;, \label{scalar}
\end{equation}
and, using (\ref{stanccr}), it is easy to show that
\begin{equation*}
<\vec{k}^{(1)},\ldots,\vec{k}^{(m)}|\vec{p}^{~(1)},\ldots,\vec{p}^{~(n)}>_{\kappa}
= \delta_{nm}\sum_{{\rm perm}(i_1,\ldots,i_n)}
\delta^{(3)}(\vec{p}^{~(1)}-\vec{k}^{(i_1)})\cdots
\delta^{(3)}(\vec{p}^{~(n)}-\vec{k}^{(i_n)})\;. 
\end{equation*}
We see therefore that the fourmonentum eigenvalues and scalar
products in $\kappa$-deformed Fock space are identical with the ones
in well-known undeformed bosonic free field theory.

We conclude that our unconventional form of $\kappa$-deformed
oscillator algebra (see (\ref{multi})-(\ref{stanccr})) is such that
applying Hopf-algebraic rules one obtains the states with standard
symmetry properties.

It appears that in our construction the notion of
$\kappa$-multiplication is very useful. It describes the
$\kappa$-deformation of the oscillators algebra as well as the
operator-valued metric defining scalar product in $\kappa$-Fock
space.

\section{Towards quantized $\kappa$-deformed field theory.}

Let us introduce the Weyl map
\begin{equation}
\vdots {\rm e}^{ip_\mu\hat{x}^\mu}\vdots \leftrightarrow {\rm
e}^{ip_\mu{x}^\mu}\;.
\end{equation}
The $\kappa$-star product, which is homomorphic to the
multiplication of two noncommutative plane waves at the same point
of $\kappa$-deformed Minkowski space \cite{lit12b}, \cite{lit12a},
looks as follows
\begin{equation}
\vdots {\rm e}^{ip_\mu\hat{x}^\mu}\vdots \cdot \vdots{\rm
e}^{iq_\mu\hat{x}^\mu}\vdots = \vdots{\rm e}^{i(p_0+q_0){\hat x}^0 +
i\Delta_i(\vec{p},\vec{q}){\hat x}^i}\vdots \leftrightarrow {\rm
e}^{ip_\mu{x}^\mu}\star {\rm e}^{iq_\mu{x}^\mu}= {\rm
e}^{i(p_0+q_0)x^0 + i\Delta_i(\vec{p},\vec{q})x^i}\;, \label{point1}
\end{equation}
where (see (\ref{c4}))
\begin{equation}
\Delta_i(\vec{p},\vec{q}) =  p_i{\rm e}^{\frac{q_0}{2\kappa}} + {\rm
e}^{-\frac{p_0}{2\kappa}}q_i\;, \label{copro}
\end{equation}
i.e. we get the Weyl map describing star product, which via
multiplication reproduces the quantum addition
law (\ref{law}). \\
We shall extend (\ref{point1}) for two different copies of
$\kappa$-Minkowski space (\ref{minkowski}) in the following way
\begin{equation}
\vdots {\rm e}^{ip_\mu\hat{x}^\mu}\vdots \cdot\vdots {\rm
e}^{iq_\mu\hat{y}^\mu}\vdots \leftrightarrow{\rm e}^{ip_\mu{x}^\mu}
\star {\rm e}^{iq_\mu{y}^\mu} = {\rm e}^{i(p_0x^0+q_0y^0) + (p_i{\rm
e}^{\frac{q_0}{2\kappa}}x^i + q_i{\rm
e}^{\frac{p_0}{2\kappa}}y^i)}\;. \label{twopoint}
\end{equation}
From (\ref{twopoint}) follows by differentiation the set of braid
relations
\begin{equation}
x_0 \star y_0 = x_0 y_0\;\;,\;\;y_0 \star x_0 = y_0
x_0\;\;\Rightarrow \;\; [\;{x}_{0},{y}_{0}\;]_{\star} =
0\;,\;\;\;\;\;\;\;\;\;\;\;\;\; \label{twocom1}
\end{equation}
\begin{equation}
x_i \star y_j = x_i y_j\;\;,\;\;y_j \star x_i = y_j
x_i\;\;\Rightarrow \;\; [\;{x}_{i},{y}_{j}\;]_{\star} =
0\;,\;\;\;\;\;\;\;\;\;\;\;\;\;\;\; \label{twocom4}
\end{equation}
\begin{equation}
x_0 \star y_i = \frac{i}{2\kappa}y_i +y_ix_0\;\;,\;\;y_i \star x_0 =
-\frac{i}{2\kappa}y_i +y_ix_0\;\;\Rightarrow \;\;
[\;{x}_{0},{y}_{i}\;]_{\star} = \frac{i}{\kappa}y_i\;,
\label{twocom2}
\end{equation}
\begin{equation}
~x_i \star y_0 = -\frac{i}{2\kappa}x_i +x_iy_0\;\;,\;\;y_0 \star x_i
= \frac{i}{2\kappa}x_i +x_iy_0\;\;\Rightarrow \;\;
[\;{y}_{0},{x}_{i}\;]_{\star} = \frac{i}{\kappa}x_i\;.
\label{twocom3}
\end{equation}
It is easy to see that if we put $x_\mu \equiv y_\mu$ we reproduce
from (\ref{twocom1})-(\ref{twocom3}) single $\kappa$-deformed
Minkowski space algebra what we consider as the basic consistency
condition.

The relations (\ref{twocom1})-(\ref{twocom3}) describe a natural
choice of braid relations between two noncommutative Minkowski
algebras with generators $\hat{x}_\mu$ and $\hat{y}_\mu$, which is
proposed for the description of the product of two fields
(\ref{field}) at different "algebraic space-time points"
$\hat{x}_\mu$, $\hat{y}_\mu$. In order to proceed with the
construction of noncommutative quantum field theory one should
consider suitable generalization of the star product and consistent
braided structures between the n copies of noncommutative
$\kappa$-Minkowski spaces. One can conjecture that the proper choice
of multilocal star product might be required (see e.g. \cite{lit21})
if we wish to derive $\kappa$-deformed Feymann diagrams with the
standard Abelian fourmomentum conservation law at the interaction
vertices\footnote{We recall that former efforts to introduce Feymann
diagrams in $\kappa$-deformed field theory (for $D=4$
$\lambda\phi^4$ model see \cite{lit18}, for $D=3$ $\lambda\phi^3$
model see \cite{lit22}) lead to the dissipation of the threemomenta
in the process of creation and subsequent annihilation of virtual
particles forming the self-energy loop.}.

The noncommutativity of the field operator has two sources -
noncommutativity of space-time and deformed quantization
of the oscillator algebra described in Sect. 2. \\
One can introduce operator-valued homomorphic Weyl map
$\phi(\hat{x}) \leftrightarrow {\hat \varphi}({x})$ with the quantum
fields ${\hat \varphi}(x)$ defined on classical Minkowski space and
depending on the quantized $\kappa$-oscillators. For bilocal
products one gets the following Weyl map
\begin{equation}
\phi(\hat{x})\phi(\hat{y})\leftrightarrow {\hat \varphi}(x)\star
{\hat \varphi}(y)\,, \label{staraa}
\end{equation}
where the operator-valued star product in (\ref{staraa}) is given by
(\ref{twopoint}). Further if we wish to  calculate the commutator
$[\,{\hat \varphi}(x),{\hat \varphi}(y)\,]_{\star}$ the relations
(\ref{multi})-(\ref{multi3}) should be used.

The application of the star product techniques and the
$\kappa$-deformed product (see e.g. (\ref{multi})-(\ref{multi3})) to
the description of quantized $\kappa$-deformed field theory is
presented in other publication \cite{our}. We recall that the
$\kappa$-deformation (\ref{a1})-(\ref{anti2}) can not be described
by twisting of classical Poincare symmetries, and the methods
implying in the case of canonical space-time noncommutativity
($\theta_{\mu\nu}$ = const) the equality of twisted and untwisted
correlation functions (see e.g. \cite{lit10}-\cite{abe},
\cite{lit23}) can not be applied.

\section{Final remarks.}

The main result of this paper was to propose new $\kappa$-dependent
quantization rules (see (\ref{multi})-(\ref{multi3})) for the field
oscillators. We show subsequently that these rules permit to
introduce the Abelian addition law of fourmomenta for
$\kappa$-deformed n-particle states, which is consistent with
coalgebra structure of $\kappa$-deformed Poincar\'{e} Hopf algebra.
The difficulty with quantum non-Abelian addition law of momenta,
which we claim it is solved in this paper, was often exposed as
argument against the physical applications of Hopf-algebraic
structure of deformed symmetries.

Unconventional feature of our proposal is use of the off-shell field
oscillators in the products (\ref{multi})-(\ref{multi3}). If we try
to consider the relations (\ref{multi})-(\ref{multi3}) with on-shell
oscillators, i.e. with the modification of energy components in a
way implying the validity of the $\kappa$-deformed mass-shell
condition, we shall face two problems:

i) In the Abelian addition law of fourmomenta (see (\ref{pstate}))
the fourmomenta $p_\mu^{(i)}$ will not satisfy the mass-shell
condition,

ii) The sum  of the on-shell energies of multiparticle states
will be not invariant under the exchange of particles.

We add here that this paper provides only basis for further
developments, which should provide the consistent and physically
plausible theory of interacting $\kappa$-deformed local quantum
fields. For that purpose one can show that our $\kappa$-deformed
oscillator algebra leads to new consistent quantization rules for
free $\kappa$-deformed quantum fields \cite{our}. One of the
problems which should be considered is the status of microcausality
condition in $\kappa$-deformed field theory. Other important problem
is the proper description of the perturbative expansion for the
interacting $\kappa$-deformed local fields. The work in these
directions is in progress.

\section*{Acknowledgements}
The authors would like to thank Piotr Kosinski for his interest in
the paper and valuable remarks.

\end{document}